# The importance of stimulus noise analysis for self-motion studies

Alessandro Nesti [1]*, Karl A Beykirch [1,2], Paul R MacNeilage [3], Michael Barnett-Cowan [1,4], Heinrich H Bülthoff [1,5]*

1: Max Planck Institute for Biological Cybernetics, Department of Human Perception, Cognition and Action, Spemannstraße 38, 72076 Tübingen, Germany

2: AMST-Systemtechnik GmbH, Lamprechthausner-Str. 63, 5282 Ranshofen, Austria

3: German Center for Vertigo and Balance Disorders, Ludwig Maximilians University Hospital of Munich, Marchioninistr. 23, D-81377 Munich

4: The Brain and Mind Institute, Department of Psychology, Natural Science Building The University of Western Ontario, London, Ontario, Canada, N6A 5B7

5: Department of Brain and Cognitive Engineering, Korea University, Seoul 136-71, Korea

*: Corresponding authors
alessandro.nesti@tuebingen.mpg.de
heinrich.buelthoff@tuebingen.mpg.de

**ABSTRACT**

Motion simulators are widely employed in basic and applied research to study the neural mechanisms of perception and action under inertial stimulations. In these studies, uncontrolled simulator-introduced noise inevitably leads to a mismatch between the reproduced motion and the trajectories meticulously designed by the experimenter, possibly resulting in undesired motion cues to the investigated system. An understanding of the simulator response to different motion commands is therefore a crucial yet often underestimated step towards the interpretation of experimental results. In this work, we developed analysis methods based on signal processing techniques to quantify the noise in the actual motion, and its deterministic and stochastic components. Our methods allow comparisons between commanded and actual motion as well as between different actual motion profiles. A specific practical example from one of our studies is used to illustrate the methodologies and their relevance, but this does not detract from its general applicability. Analyses of the simulator's inertial recordings show direction-dependent noise and nonlinearity related to the command amplitude. The Signal-to-Noise Ratio is one order of magnitude higher for the larger motion amplitudes we tested, compared to the smaller motion amplitudes. Deterministic and stochastic noise components are of similar magnitude for the weaker motions, whereas for stronger motions the deterministic component dominates the stochastic component. The effect of simulator noise on animal/human motion sensitivity is discussed. We conclude that accurate analyses of a simulator's motion are a crucial prerequisite for the investigation of uncertainty in self-motion perception.

## INTRODUCTION

For more than a century, motion simulators have been employed not only as training devices, but also in neurophysiological, psychophysical and behavioural studies that aim to inform the neural and cognitive processes of self-motion perception [1–6], as well as predicting human behaviours such as balance or aircraft control [7], [8]. In all these studies, motion trajectories executed by the simulator inevitably deviate from the commanded motion. This deviation is due to the mechanics of the device and results in motion distortions that affect amplitudes, frequencies and phases of the commanded trajectories. Throughout this paper we define total noise as the components of the actual motion that are not present in the commanded motion. We further define the total noise as the sum of a deterministic component, reproducible across repetitions of the same trajectory (e.g. mechanical deformations due to the inertia of the simulator), and a stochastic component, representing the random component of the total noise. All sensors, including the human self-motion sensory systems (visual, vestibular, auditory and somatosensory), are frequency dependent (see for example [4], [9]), and signal processing performance can be directly affected by the level of total noise in the system. Moreover, the simulator noise can provide indirect self-motion cues such as velocity dependent vibrations [10]. Response measurements such as neural, perceptual, eye movements or balance recordings should therefore be analysed with a sound understanding of the simulator capabilities and limitations, as well as the impact these limitations have on the results, so as to avoid erroneous interpretations of the data.

Surprisingly, only a few studies on human self-motion perception address the issue of simulator-introduced noise by recording the actual motion produced. Such recordings require care in minimizing sensor and environmental noise [11]. The analyses presented in these perceptual studies can be graphical and/or statistical. In a graphical analysis (see for example [12], [13]), a graphical representation of the simulator's capability is provided by plotting (in the time and/or frequency domain) the different motion recordings together. A statistical analysis, on the other hand, objectively compares recordings of different motion profiles by applying statistical tests. Note that both statistical and graphical analyses can be used to compare either two different actual motions or commanded versus actual motion. Here we summarize the main statistical approaches used so far to assess the influence of simulator noise on perceptual thresholds for self-motion and we present new methodologies for the noise analysis.

To facilitate the description of the methodologies and their relevance, we will use a psychophysical study conducted by the authors (Nesti et al., submitted). Briefly, a motion simulator was used to investigate human sensitivity to linear vertical self-motion in a range of 0 - 2 m/s$^2$. Participants were asked to discriminate a reference motion, repeated unchanged for every trial, from a comparison motion, iteratively adjusted in amplitude to measure the participants' motion discrimination thresholds. Different reference motions were tested in different experimental conditions. When interpreting the experimental results, the undesired noise introduced by the simulator is of concern for two main reasons:

1. The total noise level of reference and comparison motions within each condition, if noticeably different, would provide additional cues to the participants.
2. The increase of the total noise level with motion intensity, if non-linear, would lead to a non-constant Signal-to-Noise Ratio (SNR), resulting in differences in stimulus quality across the tested motion range.

Note that, although different experimental procedures have been proposed and used in the literature to investigate the perception of self-motion, none are immune to these problems.

The first point has been raised already by [14] and by [15]. In each of these studies, motions recorded from an inertial measurement unit (IMU) at different commanded amplitudes were analysed to assess their role in the experiments. In [14] each of their commanded trajectories was recorded multiple times (13 to 19 repetitions) and the averaged signal was subtracted from each trace to isolate the stochastic noise. Note that averaging over many repetitions causes the stochastic noise to decrease with the square root of the number of trials averaged [16], whereas the deterministic component is always present in the average signal no matter how many trials are averaged. The Fourier transform of each trace was then computed to obtain the amplitude-frequency spectrum of the stochastic noise. An ANOVA of the spectra (0.5 to 100 Hz in 0.5 Hz increments) showed no significant differences between profiles. This method provides an objective way to quantify the amount of stochastic noise in each profile by looking at the amplitude spectrum of the frequencies after the average signal is removed. Note that this procedure not only removes the commanded motion signal but also any deterministic component of the total noise. However, if there is reason to believe that the deterministic noise also depends on the motion intensity (e.g. if the amplitude of the deterministic noise increases with the amplitude of the command), the deterministic noise should not be excluded from the motion analysis, as it can provide a noticeable cue.

A different approach, employed by [15], suggests comparing two different stimuli by treating the two digital IMU measures as two different distributions after the commanded signal is filtered out in the frequency domain. A t-test between these two distributions is used to show that the amount of total noise is not significantly different. Because the t-test is specifically designed to compare the means of two populations, this method is able to detect changes in the total noise mean but remains insensitive to changes in the total noise amplitude (the distribution extremes) as long as the two signals have similar means. It is however reasonable to expect that the end-effector of the simulator oscillates around the desired trajectory yielding mean simulator noise close to zero for every trajectory. On the other hand, any effect of motion intensity on the amplitude of the noise will not be detected. . For this reason, we did not apply this methodology in the present work.

To the best of our knowledge, the second point, concerning changes in the signal quality across the tested motion intensity range, has never been addressed in any psychophysical study on self-motion perception. Even though it is known that the SNR of motion simulators depends on the commanded motion intensity (cf. [17]), the way the simulator SNR affects self-motion perception has not yet been investigated. Evidence from other fields such as object motion perception indicates that sensitivity decreases at high stimulus intensities, despite higher stimulus SNRs [18]. However with motion simulators it is difficult to obtain fine control of the SNR across the whole tested motion range. It is not our goal here to investigate the effect of the motion SNR on human self-motion sensitivity. Instead we present an SNR analysis of the motion profiles, which constitutes an essential step for a correct interpretation of experimental results.

For our chosen example study (Nesti et al., submitted) it is most appropriate to analyse the total noise, however we also present methodologies to quantify the relative contribution of stochastic and deterministic components and their dependencies on the commanded motion. Separate analysis of deterministic and stochastic components is relevant, for example, in studies where many repetitions of the same command are employed (e.g. for measuring gains of the vestibular ocular reflex). In these cases the actual motion stimulus is the motion command combined with the deterministic noise, whereas the variability across motion stimuli is determined solely by the stochastic noise.

## METHODS

The study was conducted using the Max Planck Institute CyberMotion Simulator, a 6-degrees-of-freedom anthropomorphic robot-arm, able to provide a large variety of motion stimuli, with a maximal vertical displacement of about 1.4 m and a maximal vertical linear acceleration of about 5 m/s$^2$ (for technical details refer to Robocoaster, KUKA Roboter GmbH, Germany; [19], [20]). IMU traces were acquired for 10 reference stimuli (1 Hz sinusoidal acceleration profiles with peak amplitudes of 0.07, 0.3, 1.1, 1.6 and 2 m/s$^2$, both upward and downward) with a 3D accelerometer (YEI 3-Space Sensor, 500 Hz) attached rigidly on the back of the simulator seat. Additionally, for each reference stimulus we recorded two comparison stimuli whose peak intensity was raised (higher comparison) and lowered (lower comparison) by two corresponding discrimination thresholds, so as to quantify the noise level changes within stimuli of the same condition. The discrimination thresholds associated with the reference stimuli are 0.02, 0.09, 0.21, 0.23 and 0.25 m/s$^2$, respectively (Nesti et al. submitted, unpublished observation). Each profile (Fig. 1) was recorded 20 times and low-pass filtered at 80 Hz, under the assumption that for these profiles frequencies higher than 80 Hz do not affect psychophysical performance (in agreement with [15]). From each filtered signal the 1 Hz input was subtracted to obtain the total noise signal. Fig. 2 illustrates the procedure for a downward acceleration with peak amplitude of 2.5 m/s$^2$. The deterministic component was obtained by averaging the total noise across repetitions of each profile, and the stochastic component was obtained by subtracting the deterministic component from the total noise.

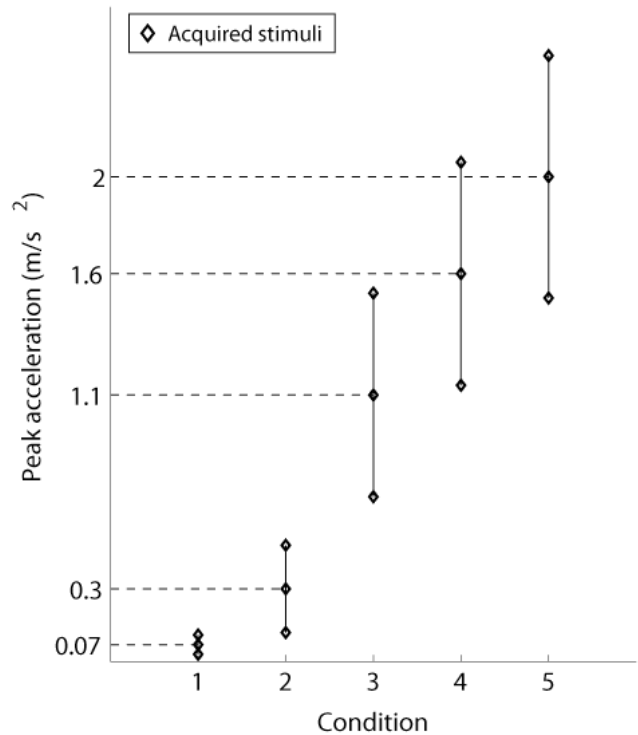

**Figure 1** *Graphical representation of the peak amplitude of the acquired stimuli, both for upward and downward motion. The dashed lines indicate the reference intensities, around which the higher and lower comparison were set*

Two different methods were employed to analyse the total noise level of the profiles: the amplitude-frequency spectrum and the root mean square (rms). These methods are explained in more details in the following sections. Additionally, for the 10 reference stimuli, the SNR was computed to characterize the relationship between the quality of the reproduced motion and the intensity of the commanded motion (section Signal-to-noise ratio analysis). We further analyse these stimuli in terms of the deterministic and stochastic components of their total noise (section Deterministic and stochastic noise analysis). Signal processing and statistical analysis were performed in MATLAB (2012a) using custom-written code and the Statistics toolbox.

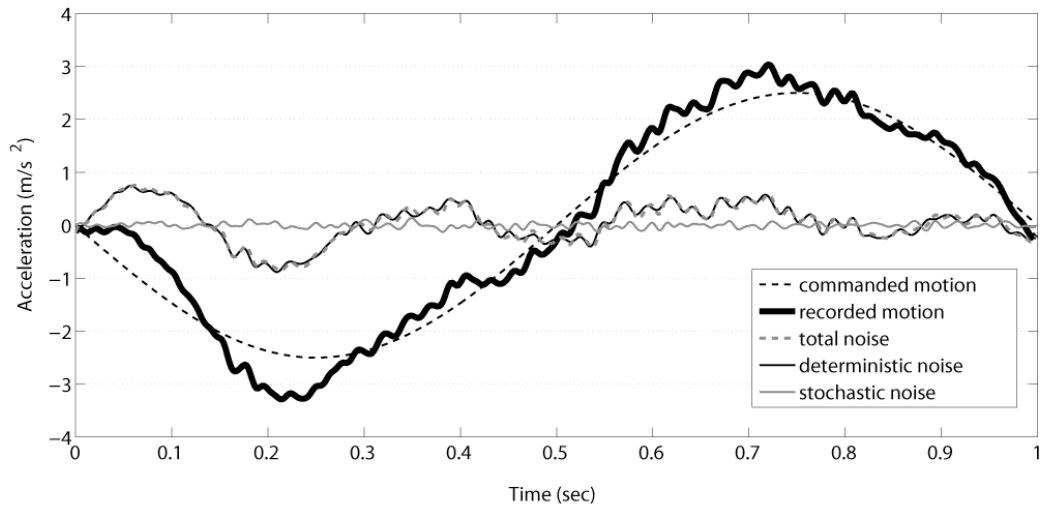

**Figure 2** *The total noise (grey dashed line) was obtained by subtracting the input command (black dashed line) from the acquired acceleration profile after low-pass filtering (black thick line). The figure also illustrates the deterministic (black thin line) and stochastic (grey thin line) components of the total noise of the recorded profile*

**Amplitude-frequency spectrum analysis**

The total noise affecting the motion profiles can be objectively quantified by its amplitude-frequency spectrum. Such an indicator has the advantage of providing details about which frequencies are more affected by the noise. This approach, based on [14], differs from the original work described previously since from each acquired trace only the input command is removed, rather than the average over several repetitions (which contains both input and deterministic noise). This allows for an analysis of the total rather than the stochastic noise. After Fourier transforming the total noise signals we obtained 3 groups of 20 amplitude-frequency spectra for each condition (Fig. 1): one group for the reference motion, one for the higher comparison and one for the lower comparison. A typical amplitude-frequency spectrum is presented in Fig. 3. Note that it is possible to infer the main frequencies that compose the total noise from spectral analysis. The 20 amplitude spectra of each reference motion were tested against the 20 amplitude spectra of their corresponding higher and lower comparisons independently by using an ANOVA with 2 factors: frequency (0 to 80 Hz in 1-Hz increments) and motion profile (reference or comparison).

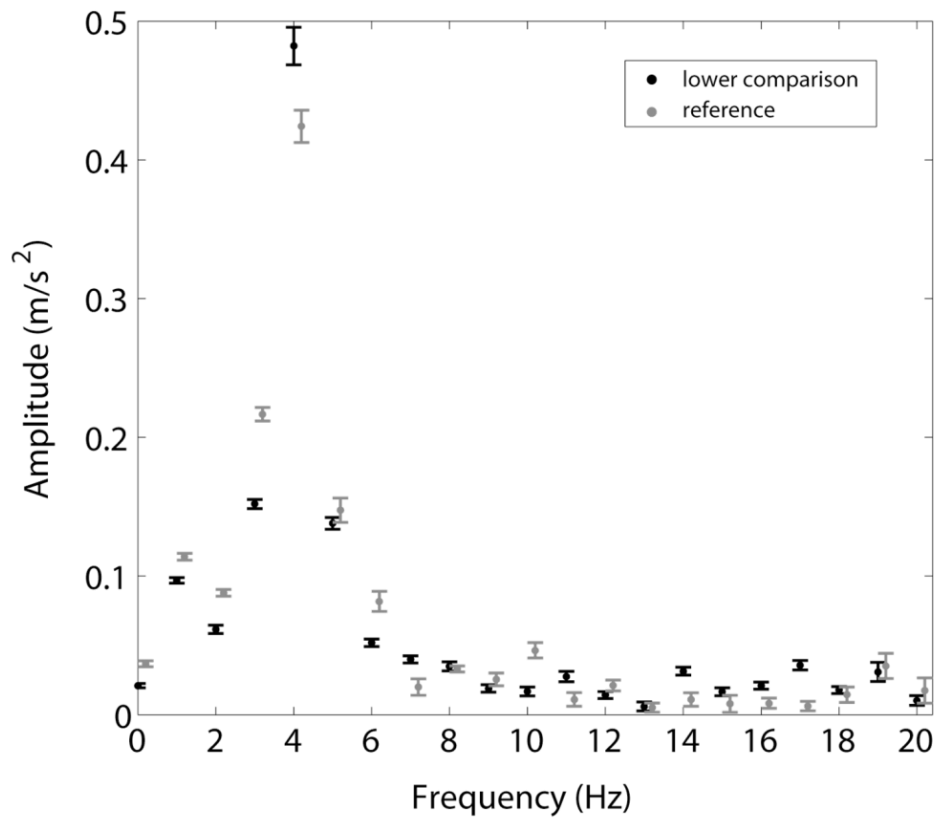

**Figure 3** *Example of an amplitude-frequency spectrum for a reference of 1.6 m/s$^2$ and corresponding lower comparison. In this example the simulator noise mainly affects frequencies around 4 Hz. Error bars represent standard deviations. The abscissa is limited to 0-20 Hz for better graphical clarity*

The same analysis was also performed on the stochastic noise so as to allow for comparison with the *analysis of vibration* reported in [14] (note the change in terminology from "vibration" to "stochastic noise"). The importance of deterministic noise is represented by the total noise analysis, since total noise is what may influence perceptual studies. The relationship of deterministic to stochastic noise is described below. Results of these analyses are shown in Table 1.

**Root mean square analysis**

The rms, or quadratic mean, is a measure of the magnitude of a varying quantity. Here, its discrete formula is used to objectively quantify the noise level of each signal:

$$rms = \sqrt{\frac{1}{N} \sum_{i=1}^{N} x_i^2} \qquad (1)$$

where $x_i$ is the i-th sampled measure of noise and $N$ is the number of samples (in our case 500 samples). For each condition we obtained 3 groups of 20 rms values each: one group for the reference motion, one for the higher comparison and one for the lower comparison. To determine whether noise level changes within conditions are reliable cues for motion amplitude discrimination, every reference rms group was tested for statistically significant differences (unpaired 2-sample t-test) against its corresponding higher and lower comparison independently. This rms analysis was conducted on the total noise as well as the stochastic component of the noise. Results of these analyses are shown in Table 1.

## Signal-to-noise ratio analysis

The SNR is used to express the relative amount of commanded signal and background noise present in each trajectory. A SNR close to 1 indicates that the level of noise in the reproduced motion is comparable to the level of commanded signal. This is often the case for motion simulators when reproducing small accelerations (e.g. <10 cm/s$^2$ for the simulator tested here) [21]. Higher SNRs indicate a reproduced motion of higher quality, where the signal level overcomes the noise. We computed the SNR for every repetition of the 10 reference motions according to the following formula:

$$SNR = \left(\frac{rms_{signal}}{rms_{total\ noise}}\right)^2 \qquad (2)$$

Differences in the SNRs were tested with an ANOVA with 2 factors: direction (upward or downward) and motion intensity (0.07, 0.3, 1.1, 1.6 and 2 m/s$^2$). See Fig. 4 for results.

## Deterministic and stochastic noise analysis

Quantitative measures of the stochastic and deterministic noise components in a reproduced motion allow for characterization of the nature of the noise, perhaps providing important information for deciding how to deal with the noise (see discussion). We calculated the rms of the stochastic and deterministic noise for the 10 reference motions. The Deterministic-to-Stochastic Ratio (DSR) introduced in equation 3 indicates which component is dominant in an analysed profile and the way that the total noise composition changes over different stimulus intensities. The results are presented in Fig. 5.

$$DSR = \frac{rms_{det\ noise}}{rms_{stoc\ noise}} \qquad (3)$$

# RESULTS

The tested reference/comparison pairs show significantly different total noise levels both in terms of amplitude-frequency spectrum and rms of the total noise signals in almost every tested pair (table 1). This indicates that the total noise introduced by the simulator depends on the commanded motion intensity.

As expected, the stochastic noise (quantified by its rms value) correlates with the inverse of the square root of the number of trials averaged (in all groups $0.77 \leq r \leq 0.99$, average $r = 0.89$). The amplitude-frequency spectrum analysis and the rms analysis were additionally performed on the stochastic noise alone. Differences in the level of stochastic noise between reference and comparison are overall smaller than for the total noise and in many cases not statistically significantly different, especially for the rms analysis (see table 1).

| Compared profiles (reference vs. comparison) [ m/s² ] | rms analysis [ t(38), p values ] | | | | Amplitude–frequency spectrum analyses [ F(1,80), p values ] | | | |
|---|---|---|---|---|---|---|---|---|
| | Total Noise | | Stochastic Noise | | Total Noise | | Stochastic Noise | |
| | Motion Direction | | Motion Direction | | Motion Direction | | Motion Direction | |
| | up | down | up | down | up | down | up | down |
| 0.07 vs. 0.03 | **<0.001** | **<0.001** | 0.50 | 0.68 | **<0.001** | **<0.001** | **<0.001** | 0.24 |
| 0.07 vs. 0.11 | **<0.001** | **<0.001** | 0.66 | 0.11 | **<0.001** | **<0.001** | 0.32 | **0.003** |
| 0.3 vs. 0.12 | **<0.001** | **<0.001** | 0.05 | **0.02** | **<0.001** | **<0.001** | **<0.001** | **<0.001** |
| 0.3 vs. 0.48 | **<0.001** | **<0.001** | 0.18 | **0.02** | **<0.001** | **<0.001** | 0.13 | **<0.001** |
| 1.1 vs. 0.68 | **<0.001** | **<0.001** | 0.39 | 0.57 | **<0.001** | **<0.001** | **<0.001** | **0.002** |
| 1.1 vs. 1.52 | **<0.001** | **<0.001** | 0.98 | 0.38 | **<0.001** | 0.08 | 0.41 | 0.05 |
| 1.6 vs. 1.14 | **<0.001** | **<0.001** | 0.36 | 0.06 | **<0.001** | **<0.001** | **<0.001** | **<0.001** |
| 1.6 vs. 2.06 | **<0.001** | **<0.001** | 0.38 | 0.19 | 0.51 | **<0.001** | **<0.001** | **<0.001** |
| 2 vs. 1.5 | **<0.001** | **<0.001** | 0.08 | **<0.001** | **<0.001** | **<0.001** | **<0.001** | **<0.001** |
| 2 vs. 2.5 | **0.039** | **<0.001** | 0.21 | **0.01** | **<0.001** | **<0.001** | **<0.001** | **<0.001** |

**Table 1** *P-values resulting from the two analyses comparing the levels of total and stochastic noise around each reference. Effects with a p-value <0.05 are considered as significant and appear in bold*

The total noise rms was found to increase non-linearly with the amplitude of the commanded signal ($F(4,199)=23562$, $p<0.001$) over the tested range, which leads to SNRs that depend on the motion intensity (Fig. 4). The results show SNRs one order of magnitude higher for the stronger than for the weaker measured profiles. Moreover, SNRs were overall better for downward compared to upward motion. These results suggest that if perceptual discrimination is modulated only by the SNR of the motion stimulus, then motion discrimination is predicted to be proportionally better for downward as compared to upward motions and for higher as compared to lower motion intensities. Experimental data on human motion sensitivity over wide motion ranges [15], [21], [22], Nesti et al. (submitted), however, contradict this prediction, suggesting an additional noise source proportional to stimulus intensity. Asymmetries in vertical motion sensitivity (Nesti et al., submitted), instead, might be entirely explained by the results of the SNR analysis.

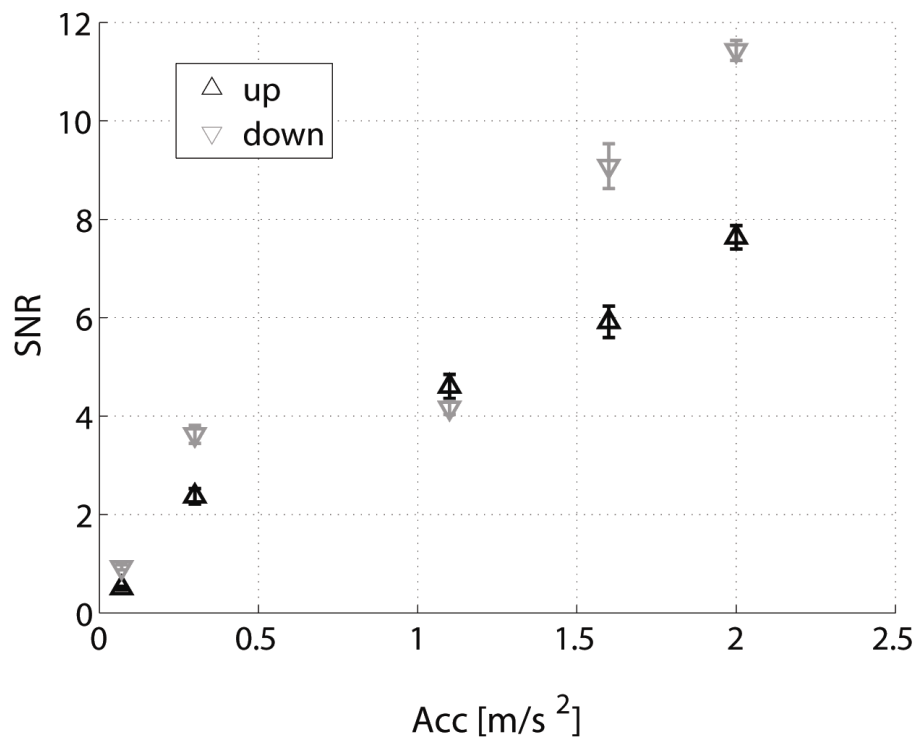

**Figure 4** *The SNR for upward and downward motion profiles increases as a function of motion intensity. Error bars represent standard deviations*

The rms of the stochastic and deterministic noise of each reference profile for upward and downward movements is presented in Fig. 5A. The level of the deterministic noise increases notably with the stimulus intensity and overall is higher than the level of the stochastic noise, which on the other hand remains rather constant over the tested motion range. For motion intensities higher than 1.1 m/s$^2$, the deterministic component of the total noise is about one order of magnitude higher than the stochastic component. The increasing DSRs (Fig. 5B) indicate that the deterministic component of the total noise dominates over the stochastic component for the stronger motions, whereas for the weaker motions the two components have the same order of magnitude. Together with the non-significant results of the stochastic noise analyses from table 1, these results suggests that deterministic noise is more likely than stochastic noise to impact self-motion perception in this experimental paradigm.

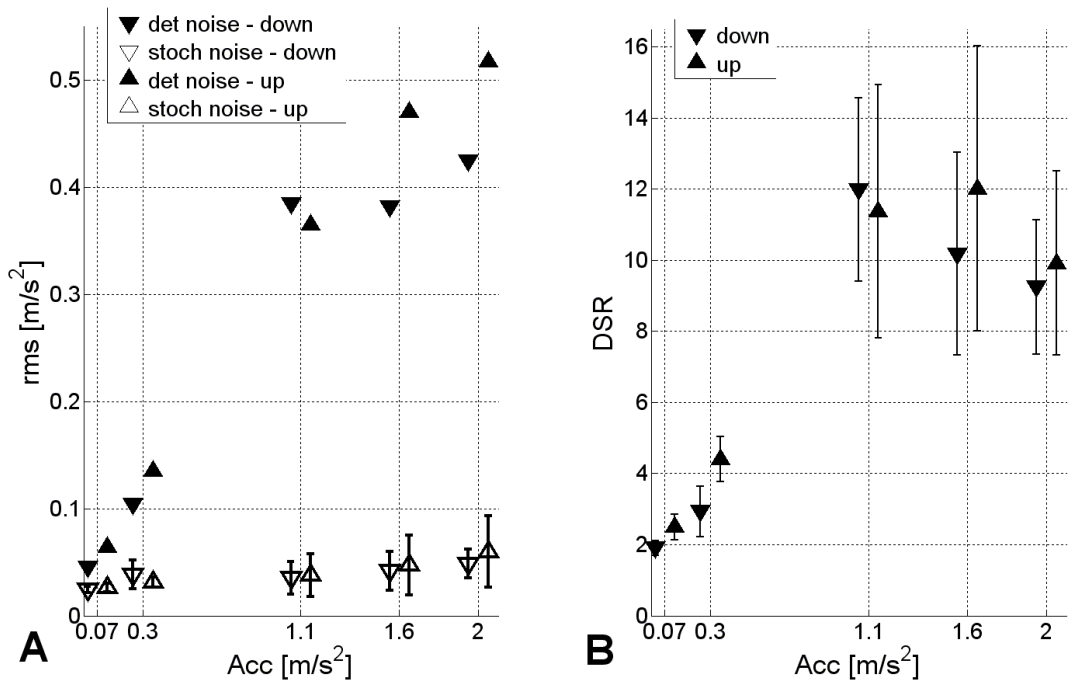

***Figure 5*** *Relative contributions of deterministic and stochastic components to the total noise.* ***Panel A:*** *Each reference stimulus recording is associated with the rms of its deterministic (filled triangles) and stochastic component (empty triangles) for both upward (upward pointing triangles) and downward (downward pointing triangles) motions.* ***Panel B:*** *The DSR for upward and downward profiles. Both panels indicate a predominance of deterministic over stochastic noise in the recorded profiles*

## DISCUSSION

The analyses presented here allow characterization of the total noise introduced by the simulator when reproducing commanded trajectories. They provide sensitive methods to compare the noise level of different commanded stimuli and to graphically and statistically describe the reproduced motion. Results show that the total noise of the simulator increases with the amplitude of the command in a nonlinear way, leading to an SNR that increases with the motion intensity. Even for relatively small changes in the amplitude of the commanded motion, changes in the measured noise are statistically significant. This raises the question of whether a human, when asked to report changes in the motion intensity, could use changes in the simulator noise as a cue rather than changes in the signal itself.

It is reasonable to assume that, if our analysis of the IMU signals do not demonstrate these differences, the human will also not detect them. It is however erroneous to conclude the opposite. The simulator motion is available to the CNS only after being processed by the sensory systems (vestibular, somatosensory and proprioceptive), whose dynamics are imperfect due to frequency dependences and noise [23], [24]. Furthermore, the way that the CNS deals with these signals may be different from the statistical analysis employed here. Consider as an example the mp3 and AAC encoding techniques in music: even though the frequency spectra of the original and compressed signals look dramatically different, they are virtually indistinguishable to a human observer due to the inability of the auditory system to perceive the differences [25]. Although the frequency response range of the otolith organs of the vestibular system is known to be between 0 Hz and 1.6 Hz [26], the contribution of the other sensory systems should not be neglected. For this reason we did not filter the data with a model of the vestibular system.

Given these results and the previous considerations, speculations can be made as to how simulator noise affects human perception of motion intensity. To distinguish between motions at different intensities, differences in the neuronal signals that reach the CNS need to overcome the internal noise level [27–29]. If the internal noise is small relative to the total noise of the motion, motion stimuli with high SNR are likely to generate neuronal signals that also have a high SNR. This would facilitate the process of detecting changes in the motion intensity. Additionally, human self-motion sensitivity could be enhanced by changes in the motion noise level if those changes are captured by the human sensors and successfully processed by the CNS. An accurate analysis of the noise of the experimental setup is therefore of great importance for the active research field investigating the noise in the nervous system and its effect on information processing [30].

In this work, an analysis of the stochastic noise alone, along with that of the total noise, is reported. The data indicate that relatively small changes in the motion input often result in significant differences in the level of stochastic noise introduced by the simulator, a difference captured more often by the amplitude-frequency spectrum analysis rather than by the rms analysis. Interestingly, different simulator behaviour is reported in [14] for a 6-degrees-of-freedom motion platform. The stochastic noise in their motion stimuli, quantified by the same analysis employed here, is reported to be not significantly different in the range of 0.1 - 0.5 $m/s^2$. This shows how different simulators can differently alter the motion command, further emphasizing the take-home message of this paper: measuring the noise of the motion stimuli is essential not only for a correct interpretation of the experimental results but also for meaningful comparisons with the existing literature.

To address the stochastic and deterministic composition of the total noise, we have provided their formal definitions and a methodology for extracting them from IMU recordings. Other than using the DSR introduced here, deterministic and stochastic noise can also be compared using a frequency analysis, which is particularly useful for highlighting the frequency ranges affected by the two types of noise (see the examples in Fig. 6).

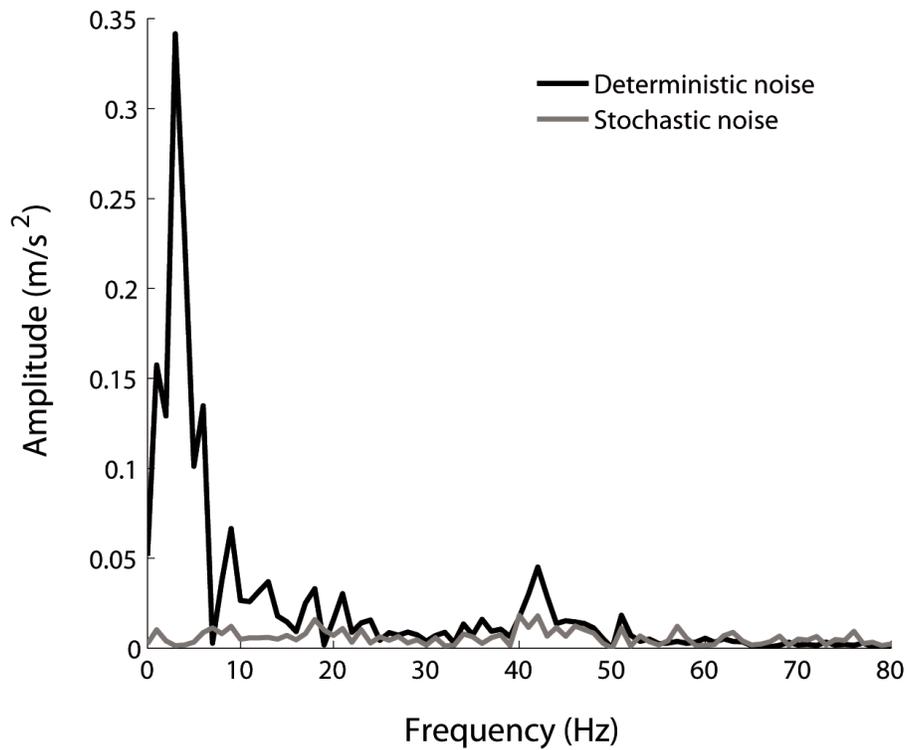

**Figure 6** *Amplitude-frequency spectrum of the deterministic and stochastic noise components of the acceleration profile whose time course is illustrated in Fig. 2. The DSR of this profile is 8.05. It is possible that the 42 Hz peak is partially environmental noise due to electric noise from the motors*

Our analysis demonstrates the importance of the total noise, which includes stochastic and deterministic components. Thus whenever possible effort should be spent in minimizing the deterministic noise, so that its impact during experiments is also minimized. This is particularly beneficial in cases where a DSR analysis indicates a predominantly deterministic nature of the noise. Deterministic noise can be reduced by using iterative learning control algorithms [31], [32]: given a desired trajectory these algorithms iteratively process IMU recordings of the simulator motion and modify the simulator commands so as to track the desired trajectory as closely as possible.

**REFERENCES**

[1] E. Mach, *Grundlinien der Lehre von den Bewegungsempfindungen*. Leipzig: W. Engelmann, 1875.

[2] R. Bárány, "Untersuchungen über den vom Vestibularapparat des Ohres reflektorisch ausgelösten rhythmischen Nystagmus und seine Begleiterscheinungen (Ein Beitrag zur Physiologie und Pathologie des Bogengangapparates)," *Monatsschr Ohrenheilk*, pp. 41477–526.526, 1907.

[3] O. Lowenstein, "The effect of galvanic polarization on the impulse discharge from sense endings in the isolated labyrinth of the thornback ray (raja clavata)," *Journal of physiology*, vol. 127, pp. 104–117, 1955.

[4] C. Fernández and J. M. Goldberg, "Physiology of peripheral neurons innervating otolith organs of the squirrel monkey. III. Response dynamics.," *Journal of neurophysiology*, vol. 39, no. 5, pp. 996–1008, Sep. 1976.

[5] A. Benson, E. Hutt, and S. Brown, "Thresholds for the perception of whole body angular movement about a vertical axis," *Aviation, Space and Environmental Medicine*, no. 60, pp. 205–213, 1989.

[6] R. Baloh, V. Honrubia, and K. Kerber, *Baloh and Honrubia's Clinical Neurophysiology of the Vestibular System*, 4th ed. Oxford University Press, 2011.

[7] R. Hosman and J. van der Vaart, "Vestibular models and thresholds of motion perception. Results of tests in a flight simulator," *Technical report LR - 265, TU Delft*, 1978.

[8] S. J. Richerson, L. W. Faulkner, C. J. Robinson, M. S. Redfern, and M. C. Purucker, "Acceleration threshold detection during short anterior and posterior perturbations on a translating platform.," *Gait & posture*, vol. 18, no. 2, pp. 11–9, Oct. 2003.

[9] G. Ellis, *Control System Design Guide, Fourth Edition: Using Your Computer to Understand and Diagnose Feedback Controllers*. Butterworth-Heinemann, 2012.

[10] S. H. Seidman, "Translational motion perception and vestiboocular responses in the absence of non-inertial cues.," *Experimental Brain Research*, vol. 184, no. 1, pp. 13–29, Jan. 2008.

[11] B. Widrow, J. R. Glover, J. M. McCool, J. Kaunitz, C. Williams, R. H. Hearn, J. R. Zeidler, E. Dong, and R. C. Goodlin, "Adaptive Noise Cancelling: Principles and Applications," vol. 63, no. 12, pp. 105–112, 1975.

[12] A. R. Naseri and P. R. Grant, "Human discrimination of translational accelerations," *Experimental Brain Research*, vol. 218, no. 3, pp. 455–64, May 2012.

[13] R. E. Roditi and B. T. Crane, "Directional asymmetries and age effects in human self-motion perception.," *Journal of the Association for Research in Otolaryngology: JARO*, vol. 13, no. 3, pp. 381–401, Jun. 2012.

[14] P. R. MacNeilage, A. H. Turner, and D. E. Angelaki, "Canal-otolith interactions and detection thresholds of linear and angular components during curved-path self-motion.," *Journal of Neurophysiology*, vol. 104, no. 2, pp. 765–73, Aug. 2010.

[15] R. M. Mallery, O. U. Olomu, R. M. Uchanski, V. a Militchin, and T. E. Hullar, "Human discrimination of rotational velocities.," *Experimental Brain Research*, vol. 204, no. 1, pp. 11–20, Jul. 2010.

[16] W. Van Drongelen, *Signal Processing for Neuroscientists*. Academic Press, 2008.

[17] Working Group AGARD, "Dynamic Characteristics for Flight Simulator Motion Systems," *AGARD Advisory Report No.144, AGARD, NATO, Neuilly sur Seine, France, September 1979.*, 1979.

[18] J. M. Zanker, "Does motion perception follow a Weber's law?," *Perception*, vol. 24, pp. 363–372, 1995.

[19] H. Teufel, H. G. Nusseck, K. A. Beykirch, J. S. Butler, M. Kerger, and H. H. Bülthoff, "MPI motion simulator: development and analysis of a novel motion simulator," in *AIAA Modeling and Simulation Technologies Conference and Exhibit*, 2007, no. 6476, pp. 1–11.

[20] M. Barnett-Cowan, T. Meilinger, M. Vidal, H. Teufel, and H. Bülthoff, "MPI CyberMotion Simulator: Implementation of a novel motion simulator to investigate path integration in three dimensions," *Journal of Visualized Experiments*, no. 63, p. e3436, 2012.

[21] L. Zaichik, V. Rodchenko, I. Rufov, Y. Yashin, and A. White, "Acceleration perception," in *AIAA Modeling and Simulation Technologies Conference and Exhibit*, 1999, no. 4334, pp. 512–520.

[22] A. Naseri and P. R. Grant, "Difference Thresholds : Measurement and Modeling," no. August, pp. 1–10, 2011.

[23] W. Gong and D. M. Merfeld, "Semicircular Canal Prosthesis," *IEEE Transaction on biomedical engineering*, vol. 49, no. 2, pp. 175–181, 2002.

[24] K. Tahboub and T. Mergner, "Biological and engineering approaches to human postural control," *Integrated Computer-Aided Engineering*, vol. 14, pp. 15–31, 2007.

[25] D. Meares, W. Kaoru, and E. Scheirer, "Report on the MPEG-2 AAC Stereo Verification Tests," 1998.

[26] W. Grant and R. Cotton, "A model for otolith dynamic response with a viscoelastic gel layer," *Journal of vestibular research*, vol. 1, pp. 139–151, 1991.

[27] S. G. Sadeghi, M. J. Chacron, M. C. Taylor, and K. E. Cullen, "Neural variability, detection thresholds, and information transmission in the vestibular system.," *The Journal of Neuroscience*, vol. 27, no. 4, pp. 771–81, Jan. 2007.

[28] J. M. Goldberg, "Afferent diversity and the organization of central vestibular pathways.," *Experimental Brain Research*, vol. 130, no. 3, pp. 277–97, Feb. 2000.

[29] X.-J. Yu, J. D. Dickman, and D. E. Angelaki, "Detection thresholds of macaque otolith afferents.," *The Journal of Neuroscience: the official journal of the Society for Neuroscience*, vol. 32, no. 24, pp. 8306–16, Jun. 2012.

[30] A. A. Faisal, L. P. J. Selen, and D. M. Wolpert, "Noise in the nervous system.," *Nature reviews. Neuroscience*, vol. 9, no. 4, pp. 292–303, Apr. 2008.

[31] H.-S. Ahn, Y. Chen, and K. L. Moore, "Iterative Learning Control: Brief Survey and Categorization," *IEEE Transactions on Systems, Man and Cybernetics, Part C (Applications and Reviews)*, vol. 37, no. 6, pp. 1099–1121, Nov. 2007.

[32] P. R. Grant, S. K. Advani, Y. Liu, and B. Haycock, "An Iterative Learning Control Algorithm for Simulator Motion System Control," in *AIAA Modeling and Simulation Technologies Conference and Exhibit*, 2007, no. August.